\documentclass[a4paper,12pt,english,german]{article}

\begin{document}

{\bf  Quantum Structures of a Model-Universe: An Inconsistency
with Everett Interpretation of Quantum Mechanics}

\bigskip

J. Jekni\' c-Dugi\' c$^{\ast}$\footnote{Email:
jjeknic@pmf.ni.ac.rs}, M. Dugi\' c$^{\dag}$, A. Francom$^{\$}$

$^{\ast}${Department of Physics, Faculty of Science, Ni\v s,
Serbia}

$^{\dag}${Department of Physics, Faculty of Science, Kragujevac,
Serbia}

$^{\$}$Austin, TX 78748, USA

\bigskip

{\bf Abstract}

We observe a Quantum Brownian Motion (QBM) Model Universe in
conjunction with recently established Entanglement Relativity and
Parallel Occurrence of Decoherence.  The Parallel Occurrence of
Decoherence establishes the simultaneous occurrence of decoherence
for two mutually irreducible structures (decomposition into
subsystems) of the total QBM model universe.  First we find that
Everett world branching for one structure excludes branching for
the alternate structure and in order to reconcile this situation
branching cannot be allowed for either of the structures
considered.  Second, we observe the non-existence of a third,
"emergent structure", that could approximate both structures and
also be allowed to branch.  Ultimately we find unless
world-branching requires additional criteria or conditions, or
there is a privileged structure, that we provide a valid model
that cannot be properly described by the Everett Interpretation of
Quantum Mechanics.

\bigskip


\pagebreak

{\bf 1 Introduction}

\bigskip

\noindent
  {\it  "Today, it is often said that in Everettian quantum theory the notion of
    parallel 'worlds' or 'universes' applies only to the macroscopic worlds defined
    (approximately) by decoherence. Formerly, it was common to assert the existence
    of many worlds at the microscopic level as well. Without entering into any
    controversy that might still remain about this, here for completeness we shall
    address the Claim for both 'microscopic' and 'macroscopic'
    cases."} ( Valentini 2010)

    \smallskip

As the Everettian quantum theory claims to be a valid
interpretation of the universally valid quantum mechanics, there
should be no known model in which it fails to perform.

Using the Quantum Brownian Motion (QBM) model (Giulini et al 1996;
Breuer and Petruccione 2002; Caldeira and Leggett 1985; Romero and
Paz 1997) as a "model universe", Entanglement Relativity (ER)
(Dugi\' c 1999; Dugi\' c and Jekni\' c 2006; Ciancio et al 2006;
Dugi\' c and Jekni\' c-Dugi\' c 2008; de la Torre et al 2010;
Harshman and Wickramasekara 2007; Jekni\' c-Dugi\' c and Dugi\' c
2008; Terra Cunha et al 2007; Tommasini et al 1998), and the
Parallel Occurrence of Decoherence (POD) (Dugi\' c and Jekni\'
c-Dugi\' c 2012), we will demonstrate for the first time using
these methods, that Everett interpretation fails to properly
describe a physically relevant decoherence model.

The QBM model affords us the opportunity to avoid getting involved
with extending or arguing in the realm of interpretation because
firstly it is a non-trivial model universe (a model of an isolated
quantum mechanical system) referring to a realistic and
mathematically well-defined physical situation, and secondly, the
QBM model is paradigmatic for the Everett Interpretation ( S.
Saunders 2010; Wallace 2010; Hartle 2010): as it directly
implements the quasi-classical dynamics of the
decoherence-selected basis of wavepackets approximately localized
in both position and momentum of the open system (of the Brownian
particle).

We consider a pair of structures (decompositions into subsystems),
formally denoted $1+2$ and $S'+E'$, of the model universe. The
structures are mutually related by the proper Linear canonical
transformations (LCTs). This kind of restructuring the Universe is
not new, see e.g. Saunders (2010), and Kent (2012). However, our
focus goes beyond these general considerations in that it is
devoted to the mutually {\it irreducible} structures, $1+2$ and
$S'+E'$. This irreducibility of the structures provides the first
main point of our consideration: For the model considered, the
Everett branching for one structure excludes Everett branching for
the other structure. In order to reconcile the two, we conclude
that Everett branching cannot take place for any of the
structures.

Modern Everett interpretation assumes there need not be branching
for the "microscopic" model of a composite system. Rather, some
effective, "emergent" structure (describable by the "higher level
ontology") is expected to branch Wallace (2010, 2012). So, our
first finding may not be relevant.

However, for the QBM model,  the second main point of our
considerations reads: we do not find physical degrees of freedom
that might support the emergentism of modern Everett
interpretation. Thereby we are forced to conclude that the Everett
interpretation cannot properly describe the QBM universe-model.

A possible way out of this inconsistency with Everett theory might
be to choose one structure as physically relevant (preferred), at
the expense of rejecting the alternate structure as physically
irrelevant, artificial.

In the absence of  a physically sound and clear criterion for
making the choice of the preferred universe structure, we finally
conclude that there is a physically relevant model that cannot be
properly described by the Everett interpretation of quantum
mechanics.

In Section 2 we introduce the concept of entanglement relativity.
In Section 3 we briefly present the recently obtained parallel
occurrence of decoherence for  quantum Brownian Motion (Dugi\' c
and Jekni\' c-Dugi\' c 2012). Section 4 provides the main result
of our paper. Section 5 is discussion and we conslude in Section
6.

\bigskip

{\bf 2.  Entanglement relativity}

\bigskip

Below, we will discuss Entanglement Relativity (ER) as a subtle,
and perhaps often overlooked aspect of the universally valid
quantum mechanics. With an eye towards this subtlety, references
(Dugi\' c 1999; Dugi\' c and Jekni\' c 2006; Ciancio et al 2006;
Dugi\' c and Jekni\' c-Dugi\' c 2008; de la Torre et al 2010;
Harshman and Wickramasekara 2007; Jekni\' c-Dugi\' c and Dugi\' c
2008; Terra Cunha et al 2007; Tommasini et al 1998) demonstrate
how ER appears in quantum mechanics related articles. For the sake
of clarity we will now highlight and analyze its important
aspects.

The hydrogen atom is defined as a bipartite system "electron +
proton ($e + p$)". However, in practice it is recognized as a pair
"center of mass + relative coordinates ($CM +R$)". These two
atomic decompositions (structures) are mutually linked by the well
known linear canonical transformations (LCTs) which introduce $CM$
and $R$ to the atom.  The relevant LCTs allow for the "separation
of variables" and for the exact solution to the Schr\" odinger
equation in the $CM + R$ decomposition.  The related quantum state
(while neglecting the atomic spin) is in tensor-product form
$\vert \chi \rangle_{CM} \vert nlm_l\rangle_R$, where $n, l, m_l$
are the well known numbers from quantum theory for the hydrogen
atom.

On the other hand, the Coulomb interaction between $e$ and $p$
leads to the conclusion that the pair $e + p$ must be in an
entangled state of the form $\sum_i c_i \vert i \rangle_e \vert i
\rangle_p$. Bearing in mind that $e + p =$atom$= CM + R$, the
universally valid quantum mechanics implies the following equality
(for an instant of time):

\begin{equation}
\sum_i c_i \vert i \rangle_e \vert i \rangle_p = \vert
\chi\rangle_{CM} \vert n l m_l  \rangle_R.
\end{equation}

\noindent Dynamically: the $R$-system's state is stationary
(multiplied by $\exp(-\imath t E_n/\hbar)$), while the
$CM$-system's state may be e.g. a wave packet freely evolving in
time.

Eq. (1) is paradigmatic for ER in that a change of the spatial
degrees of freedom of a composite system typically results in a
formal change of the composite system's quantum state. There is
entanglement for every instantaneous quantum state of a composite
quantum system and the very concept of entanglement is {\it
relative}.

In a more abstract form eq. (1) reads as follows, where, of
course, $1 + 2 = C = S + S'$:

\begin{equation}
\sum_i c_i \vert i \rangle_1 \vert i \rangle_2 = \vert
\chi\rangle_S \vert \psi \rangle_{S'},
\end{equation}

Then, one may undertake the task of kinematically transforming the
left hand side into the right hand side of eq. (2),  and {\it vice
versa}. Generally, this is a formidable task not yet very well
known. However, for some models (Breuer and Petruccione 2002;
Caldeira and Leggett 1985; Romero and Paz 1997; Dugi\' c and
Jekni\' c 2006; de la Torre et al 2010), we already know about the
validity of eq. (2) as a corollary of the universally valid
quantum mechanics.

Equation (2) applies to a system ${C}$ of arbitrary complexity. To
illustrate, one can first directly generalize ER as follows:

\begin{equation}
\sum_p \alpha_p \vert \varphi_p \rangle_1 \vert \Phi_p \rangle_2 =
\sum_{k} \beta_k \vert \mu_k\rangle_S \vert \nu_k \rangle_{S'}.
\end{equation}

Now, two remarks are in order regarding eq. (3).  First, every
subsystem of ${C}$ ($1$; $2$; $S$; $S'$) may bear its own
structure and related internal entanglement.  Second, the above
expressions equally refer to the cases when certain subsystems
(e.g. the atomic spin, or the system's environment) are neglected
or implicit or that are not yet known. Thereby ER eq. (2) equally
addresses the hydrogen atom as well as the quantum Universe.  For
the hydrogen atom in a non-relativistic frame eq. (2) is precise,
while for "the Universe" the expression of eq. (2), and likewise
of eq. (3), assumes that further decomposing of universal
subsystems is possible.

Eq. (2) reveals the presence of nontrivial non-negligible quantum
entanglement in a composite quantum system relative to the
pertinent degrees of freedom.  We do not account for the change in
entanglement due to a change of the reference frame (Gingrich and
Adami 2002). ER otherwise is effectively ubiquitous.

Entanglement Relativity therefore is a descriptive name for that
there is always an entangled form of a composite system's state.
In its kinematical form eq. (2), ER directly points out
inconsistency in the original Everett's "relative states"
interpretation Everett (1957). A related dynamical consideration
easily points out inconsistency in the modern Everett
interpretation. The arguments will be given in Section 4.

\bigskip

\bigskip

{\bf 3. Parallel occurrence of decoherence for irreducible
structures of the model universe }

\bigskip

The occurrence of decoherence for the Universe is a plausible
conjecture, an extrapolation of the existing (typically very
simple) models of the decoherence theory. While there is progress
in describing  ever more-complex systems by decoherence, the truly
complex systems, such as the Universe, are as yet out of reach.
Therefore there is no alternative to modeling the Universe except
but to employ relatively simple models.

Nevertheless, considering simple models does not decrease the
importance of the conclusions obtained as the Everett
Interpretation has pled to full universality, that is, to
model-independence. In principle, this makes the simple models
equally useful for drawing general conclusions within the Everett
Interpretation, cf. e.g. Barvinsky and Kamenshchik (1995).

Bearing that in mind, we examine the well-known Quantum Brownian
Motion (QBM)  model (Giulini et al 1996; Breuer and Petruccione
2002; Caldeira and Leggett 1985; Romero and Paz 1997). First, it
is a nontrivial decoherence model referring to a realistic
physical situation. Second, the QBM model directly distinguishes
the Gaussian states as the decoherence-selected "preferred states"
of paramount importance for the Everett Interpretation:

\smallskip

\noindent{\it " In contrast, states well localized in phase
space--wavepackets--reliably decohere, and even though elements of
a superposition, evolve autonomously from each other for a wide
class of Hamiltonians. With respect to states like these,
Ehrenfest's theorem takes on a greatly strengthened form. But
decoherence in this sense is invariably approximate; it is never
an all-or-nothing thing."} (Saunders 2010)

\bigskip

{\bf 3.1 Definition of the structures. Irreducibility}

\bigskip

Consider a pair of systems, $1$ and $2$. The $1$ system is one
dimensional, i.e. described by the position and the momentum
observables, $x_1$ and $p_1$, respectively; the commutator, $[x_1,
p_1] = \imath \hbar$. The $2$ system is many-particle. The
constituent particles described by the respective position and
momentum observables, $x_{2i}$ and $p_{2i}$; with the commutator
$[x_{2i}, p_{2j}] = \delta_{ij} \imath \hbar, i,j=1,2,...N$. The
composite system, $C$, is defined, $C=1+2$.

Now we introduce alternate structure of $C$. We do this by
employing the proper LCTs that introduce a new pair of systems,
$S'$ and $E'$. with the respective conjugate observables, $
X_{S'}$ and $P_{S'}$, and $\rho_{E'\alpha}$ and $p_{E'\alpha}$.
Again, we assume the new $S'$ system is one-dimensional. Of
course, $1+2 = C = S'+E'$.

Every observable $x_1, x_{2i}$ is assumed to be a linear
combination of the alternate observables, $X_{S'}$ and
$\rho_{E'\alpha}$, and {\it vice versa}; e.g., $X_{S'} =
\sum_{i=1}^{N+1} c_{i} x_i$, $\rho_{E'\alpha} = \sum_i b_{\alpha
i} x_i, \alpha=1,2,...,N$. This makes the two structures, $1+2$
and $S'+E'$, mutually {\it irreducible}.

The irreducibility means (for some details see Section 4.2): a)
that the $S'$ (likewise the $E'$) system cannot be decomposed or
partitioned in to the $1$ and $2$ system and {\it vice versa}, and
furthermore b) the degree of freedom of the $1$ system (of the
$S'$ system), is a linear combination of all of the degrees of
freedom of the  $S'$ and $E'$ (of the $1$ and $2$) systems.
Consequently no measurement of the $1$  system represents a
measurement of the  $S'$ system, and {\it vice versa}. Finally, c)
the probability density of one subsystem does not yield a
probability density for any other subsystem of the alternate
structure.

The points (a)-(c) directly give rise to: (i) by very definition,
every subsystem, $1, 2, S', E'$, belongs to only one structure,
either to $1+2$ or to $S'+E'$; (ii) by very definition, both the
$1$ and $S'$ systems are one-dimensional and cannot exchange
particles with the rest, $2$ and $E'$; (iii) due to (b) and (c), a
local observer [belonging either to $1+2$ or to $S'+E'$] cannot
obtain any information about the $S'$ system based on the
information about the $1$ system, and {\it vice versa}; (iv) due
to (b) and (c), there is not any information flow between the
subsystems belonging to the different structures.

Of course, we do not claim universal validity of the above
(a)-(c), i.e. of (i)-(iv). There easily can be constructed
mutually reducible structures for which some of these need not
apply.

\bigskip

{\bf 3.2 Parallel decoherence for the QBM model}

\bigskip

We briefly present the results of Dugi\'c and Jekni\' c-Dugi\' c
(2012). We borrow notation from Section 3.1. The QBM model
considers a point-like particle $1$ (or the particle's
center-of-mass) interacting with  harmonic-bath oscillators
(system $2$). The composite system ${C} = 1+2$ is defined by  the
${C}$'s state-space $H_1 \otimes H_2$ tensor-product structure and
the total Hamiltonian:

\begin{eqnarray}
&\nonumber&  H = { p_1^2 \over 2m_1} + V( x_1) + \sum_i ({
p_{2i}^2 \over 2 m_{2i}} + {m_{2i} \omega_i^2  x_{2i}^2 \over 2})
\\&& \pm  x_1 \sum_i \kappa_i  x_{2i} \equiv  H_1 +
 H_2 +  H_{1+2},
\end{eqnarray}

\noindent where the index $i$ enumerates the environmental
particles, and the sign $\pm$ is in accordance with the variations
of the model contained in the literature. The physically relevant
open system models are usually considered (cf. e.g. Breuer and
Petruccione 2002): $V( x_1) = 0$ for the free particle, or $V(
x_1) = m_1\omega^2 x_1^2/2$ for the harmonic oscillator.

The initial state $ \rho_{C}$ of the pair $1+2$ is separable, $
\rho_{C} =  \rho_1 \otimes  \rho_{2th}$, while $ \rho_{2th}$ means
that the harmonic-bath environment is in thermal equilibrium.  The
general QBM theory (Giulini et al 1996; Breuer and Petruccione
2002; Caldeira and Leggett 1985; Romero and Paz 1997) states: The
open system $1$ is subject to decoherence induced by its
environment $2$, while related "pointer basis" (the robust,
quasi-classical) states are the Gaussian states. The Gaussian
state dynamics is very much like that which would be expected  in
a classical system: due to the large environment $2$, decoherence
effectively irreversibly destroys the linear superpositions of the
Gaussian states of the system $1$; the environment $2$ effectively
performs the approximate position-measurement for system $1$.
System 1, the "Brownian particle", exhibits quasi-classical
behavior very much like the "classical Brownian particle".
Therefore the composite ${C}$ system is a proper model-universe.

Now, we introduce another structure, $S' + E'$, for the isolated
composite system ${C}$ (Dugi\' c and Jekni\' c-Dugi\' c 2012). The
Hilbert state space of $C$ is now factorized, $H = H_{S'} \otimes
H_{E'}$. We first introduce the standard $CM$ (center of mass) and
$R$ (relative positions) for the total system: $X_{CM} = \sum_i
m_i x_i/\sum_j m_j$ and $r_{R\alpha} = x_i - x_j$ , $\alpha \equiv
(i, j)$, and $P_{CM}$ and $p_{R\alpha}$ represent the respective
conjugate
 momentums.
 The Hilbert state space factorizes $H = H_{CM} \otimes H_R$.

 For the $H_1 + H_2$ part of the Hamiltonian eq. (4),
 i.e for the set of non-interacting particles, it is known (Mc Weeny 1978) that $H_1 + H_2$
 transforms as follows. The kinetic terms for all the new
 particles takes the standard form $p^2/2m$; the $CM$ system's
 mass is the total $S+E$ system's mass, $M$, while the $R$ system's constituent
 oscillators'
 masses are the reduced masses, $\mu_{\alpha}$ for the $\alpha$th
 oscillator. There appears the so-called "mass polarization" term
 that bi-linearly couples the momentums of the environmental oscillators.
Physically, the new structure $CM+R$ consists of the  $CM$ system,
which does not interact with the $R$ system,
 and the $R$ system is a set of linearly coupled oscillators.

 When we take the interaction $H_{1+2}$ into account, one
 easily obtains: there appear additional quadratic terms (harmonic
 potentials) for the $CM$ system as well as for every $R$-system's
 oscillator, while there is a linear coupling between the position
 observables for the $CM$ and $R$-system's oscillators, of the
 same form as given in eq. (4). So, the $CM$ system is a harmonic
 oscillator even if it was not the case for the original open
 system $1$. There also appear bi-linear coupling of the position
 observables for the environmental oscillators. In effect, there
 is bilinear coupling of the position, as well as of the momentum
 observables for the environmental (for the $R$'s) oscillators.

For the oscillators bi-linearly coupled via their position and/or
momentums, one can always perform another linear
variables-transformation in order to decouple the oscillators (Mc
Weeny 1978). So, for the $R$ system we can introduce the new
position observables (the normal modes), $\rho_{R\alpha}$, and the
related conjugate momentums, $\pi_{R\alpha}$, for which the
$R$-system becomes a set of mutually uncoupled oscillators. With
the notation, $S' \equiv CM$ and $E' \equiv R$, the Hamiltonian
eq. (4) acquires the form:

\begin{eqnarray}
&\nonumber&  H = { P_{S'}^2 \over 2M} + {1 \over 2} M \Omega^2
 X_{S'}^2+ \sum_{\alpha} ({ \pi_{E'\alpha}^2 \over 2
\mu_{\alpha}} + {1 \over 2} \mu_{\alpha} \nu_{\alpha}^2
\rho_{E'\alpha}^2)  \\&& \pm  X_{S'} \sum_{\alpha} \sigma_{\alpha}
\rho_{E'\alpha} \equiv  H_{S'} +  H_{E'} + H_{S'+E'}.
\end{eqnarray}

\noindent which is formally  isomorphic with the form of eq. (4)
for the original $1+2$ structure.

There are a number of details differentiating  between the two
structures, $1+2$ and $S'+E'$ (Dugi\' c and Jekni\' c-Dugi\' c
2012, Jekni\' c-Dugi\' c et al 2013). Nevertheless, as it is
well-known (Giulini et al 1996; Breuer and Petruccione 2002;
Caldeira and Leggett 1985; Romero and Paz 1997), the occurrence of
the Brownian effect is largely independent from the details
distinguishing both models, eq. (4) and eq. (5). Particularly, the
occurrence of decoherence (of the effective approximate
position-measurement of the Brownian particle) is independent of
the presence/absence of correlations (quantum or classical) in the
initial state, of the strength of the interaction in the composite
system "Brownian particle + harmonic-oscillators-environment", or
on the so-called form of the "spectral density". The formal
similarity between the two models, eq. (4) and eq. (5), allows the
following conclusion on the {\it parallel occurrence of
decoherence} (POD) (Dugi\' c and Jekni\' c-Dugi\' c 2012):

The unitary evolution of the initial state $ \rho_C$ generated by
the Hamiltonian $ H$ gives, for the different structures of ${C}$:
For as much as System $1$ represents the "Brownian Particle", in
the $1+2$ decomposition, System $S'$ represents the "Brownian
Particle" for the $S'+E'$ decomposition.

Physically, both structures bear the decoherence-induced
quasiclassicality that is required in the foundations of modern
Everett interpretation. So we have a model Universe $C$ with the
two structures that are simultaneously and quasiclassically
evolving in time, $1+2$ and $S'+E'$.

\bigskip

{\bf 4. The inconsistency}

\bigskip

It is worth repeating: Both structures, eq. (4) and eq. (5), are
formally equal and mutually irreducible. Both structures bear a
physically clear "system-environment split" with  large
environments capable of providing  fast decoherence as an
effectively irreversible process. As both Brownian particles, $1$
and $S'$, are elementary (one-dimensional), the number of degrees
of freedom in $2$ is equal to the number of the degrees of freedom
in the $E'$ environment. The initial state for both decompositions
is the same, while being subject to ER, Section 2. The two
decoherence processes (for $1$ and for $S'$) unfold simultaneously
in time. The basis (the "preferred states") picked out by
decoherence {\it for both open systems} is approximately e.g a
"coherent-state" (a wavepacket) basis whose dynamics are
quasi-classical in the sense that the behavior of those
wavepackets approximates the behavior predicted for the classical
Brownian particle.

\bigskip

{\bf 4.1 Non-branching of the structures}

\bigskip

Let us  consider the dynamics of the $1+2$ structure.  The
decoherence-preferred states are known to be Gaussians (Giulini et
al 1996; Breuer and Petruccione 2002; Caldeira and Leggett 1985;
Romero and Paz 1997). Then one can write (cf. eq. (5) in Wallace
2010) for the universal state:

\begin{equation}
\vert \Psi (t) \rangle_C = \int d x_1 d p_1 \alpha (x_1, p_1, t)
\vert x_1, p_1 \rangle_1 \vert \epsilon (x_1, p_1)\rangle_2,
\end{equation}

\noindent where $\vert x_1, p_1 \rangle_1 $ is a "coherent state"
for the $1$ system in correlation with the approximately
orthogonal environmental states, $\vert \epsilon(x_1,
p_1)\rangle_2$. In the presence of decoherence, Section 3.2,
$\vert\alpha (x_1, p_1, t)\vert^2$ evolves, to a good
approximation, like a classical probability density on phase space
for the $1$ open system. Owing to the correlations in Eq. (6), one
can define a set of consistent histories for the total system
$1+2$. One such history for the time instants $t_{\circ} < t_1 <
t_2...$ :

\begin{eqnarray}
&\nonumber& \vert x_1(t_{\circ}), p_1(t_{\circ}) \rangle_1\vert
\epsilon(x_1(t_{\circ}), p_1(t_{\circ}))\rangle_2 \to \vert
x_1(t_1), p_1(t_1) \rangle_1\vert \epsilon(x_1(t_1)),
p_1(t_1))\rangle_2
\\&&
\to \vert x_1(t_2), p_1(t_2)) \rangle_1\vert \epsilon(x_1(t_2),
p_1(t_2))\rangle_2...
\end{eqnarray}

\noindent [defined with some probability], approximately
represents  dynamics of an Everet world.

The point to be strongly emphasized: Entanglement Relativity,
Section 2, refers to {\it every} instant in time. A tensor-product
state $\vert x_1(t_i), p_1(t_i) \rangle_1\vert \epsilon(x_1(t_i),
p_1(t_i))\rangle_2$ obtains {\it entangled form} for the alternate
structure $S'+E'$ for practically every time instant $t_i, i = 0,
1, 2,...$. As a consequence, the set of mutually consistent
histories for the $1+2$ structure is not "consistent" for the
structure $S'+E'$. So Everett branching for the $1+2$ structure
{\it excludes} the Everett branching for the $S'+E'$ structure.

Section 3.2 now suggests the roles of the two structures in this
analysis can be inverted. More precisely, as the two structures
are equally valid decompositions of the Universe, exchanging the
roles of the structures in the above analysis leads us to the
following observation: Everett branching for the $S'+E'$ structure
{\it excludes} Everett branching for the $1+2$ structure.

In effect, an Everett branch (an Everett world) for one structure
cannot last longer than the decoherence-induced branching for the
alternate structure. Bearing in mind the fact that decoherence is
a fast process, also for the model considered in Section 3.2
(Dugi\' c and Jekni\' c-Dugi\' c 2012), we find:  Mutually
exclusive yet simultaneous splitting processes for the two
structures effectively result in the impossibility of
World-Branching for both structures.

It is important to emphasize: Everett branching is not in conflict
with either   ER or  POD, separately, it is {\it not consistent}
with ER {\it and} POD together. On the other hand, both ER and POD
are  corollaries of the universally valid quantum mechanics.
Therefore we conclude:

\noindent {\it The main interpretational rule of branching of the
Everett worlds is in conflict with the universally valid quantum
mechanics for the QBM model of Section 3}.

\bigskip

{\bf 4.2 In search of emergent structure for the QBM model}

\bigskip

Decoherence is typically studied starting from a fairly
unprincipled choice of system-environment split. In this sense,
decoherence is by its nature an approximate process and so the
process of branching is likewise approximate. In other words
(Wallace 2010) [our emphasis]: "...decoherence is an {\it
emergent} process occurring within an already stated microphysics:
unitary quantum mechanics. It is not a mechanism to define a part
of that microphysics". This is the basis for the modern Everett
interpretation that can be expressed as follows [our emphasis]:

\noindent "{\it There is just a dynamical
process--decoherence--whereby certain components of that state
become dynamically autonomous of one another. Put another way: if
each decoherent history is an} emergent {\it structure within the
underlying microphysics, and if the underlying microphysics
doesn't do anything to prioritize one history over another (which
it doesn't) then all the histories exist. That is: a unitary
quantum theory with} emergent, {\it decoherence-defined
quasiclassical histories is} a many-worlds theory." (Wallace 2010)

Within this new wisdom, one may suppose that there should be an
emergent structure for the model-Universe $C$ that should branch.
In other words: the structures $1+2$ and $S'+E'$ may be considered
"microscopic" i.e. of no interest individually for branching. This
can make the finding of Section 4.1 pointless.

To test whether the finding of Section 4.1 is pointless in this
light, one should construct an emergent model-structure eligible
for the branching.

In the absence of a general physical definition of "emergent
properties" (i.e. of the "higher level ontology") of complex
systems, cf. e.g. Auyang (1998), we are forced to speculate about
the possible ways to obtain a branching-eligible structure for the
QBM model of Section 3. To this end, we are able to detect only
two possibilities. We find both of them inappropriate for defining
an emergent QBM structure or for recognizing their decoherence
induced dynamics to approximate each other.

We distinguish the following bases for emergentism. {\it First},
it is the dynamical exchange of particles between the "system" and
the "environment" that encompasses the standard choice of the
"dividing line"  in the von Neumann sense (the von Neumann
"chain"), von Neumann (1955). {\it Second}, one may suppose there
is an alternate, third structure providing an emergent Brownian
particle, $B$, for the pair of Brownian particles, $1$ and $S'$,
Section 3.

To see that the first doesn't work for the QBM model is
straightforward. Actually, both Brownian particles are
one-dimensional and there is not, {\it by definition}, any
possibility of exchanging particles of e.g. the  $1$ system with
the environment $2$ (or of the $S'$ system with the environment
$E'$); of course, {\it due to irreducibility} of the two
structures, exchange of the particles between the $1$ system with
the environment $E'$ (i.e. of the $S'$ system with $2$) is not
even defined. The variant that an environmental oscillator takes
the role of Brownian particle is in principle also not allowed. In
both models, Eq. (4) and Eq. (5), the environmental particles do
not mutually interact and therefore there is not a properly
defined environment for the variant--not  to mention that this
excludes the possibility (Giulini et al 1996; Breuer and
Petruccione 2002; Caldeira and Leggett 1985; Romero and Paz 1997)
that the $1$ system is a "free particle" (not an oscillator). So
the system-environment split cannot be altered for the QBM model.

The second option is a bit more subtle yet. To this end we justify
the claims of Section 3.1: (1) obtaining information about one
Brownian particle (e.g. $1$) provides no information about the
alternate one (e.g. the particle $S'$); (2) there does not exist
any observable, $X_B$, of the subsystem $B$ of the composite
system $C$ that would approximate measurements of any pair of
observables of the two Brownian particles, $1$ and $S'$. In
effect, there is not any structure $B+E_B$ of $C$ that could be
emergent {\it for} the structures $1+2$ and $S'+E'$.

Regarding the  point (1), we first remind (cf. Section 3.1): the
variables of the $1$ (of the $S'$) are linear functions of all the
variables of the $S'$ and $E'$ (of the $1$ and $2$) systems. So
local measurements performed on the $S'$ system reveals nothing
about the $1$ system, and {\it vice versa}. More formally: $tr_2
\vert \Psi\rangle\langle \Psi\vert$ provides a probability
density, $\rho(x_1, x'_1)$, for the $1$ system, and analogously
for the $S'$ system. It is obvious that $\rho(x_1, x'_1)$ {\it
cannot in principle} be used to define any probability density for
$S'$, and {\it vice versa}. e.g. The integral $\int \rho(x_1,
x'_1) \Pi^{\otimes \alpha} d\rho_{E' \alpha}$ is {\it not} a
probability density for the $S'$ system. Furthermore, linear
dependence of the observables $x_1$ and $X_{S'}$ makes the
"tracing out" e.g. of the form $\int \rho(x_1, X_{S'}) dX_{S'}$
ill-defined. Thereby one can say the Brownian particles, $1$ and
$S'$, belonging to the mutually irreducible structures, are
mutually information-theoretically separated.  The two decoherence
processes, for the $1$ system and for the $S'$ system, cannot
approximate each other, nor is there any information flow between
$1$ and $S'$. The conclusion refers to every subsystem, including
the observer of a structure. An observer belonging to the $1+2$
structure is a subsystem of the $2$ environment, while the
observer belonging to the $S'+E'$ structure is a subsystem of the
$E'$ environment. Then e.g. by exchanging the $x_1$ above by the
observer's position, $x_{obs}$, we conclude: the alternative
structure's (the $S'+E'$ structure's) subsystems are
information-theoretically separated from the observer belonging to
the alternate (i.e. to the $1+2$) structure--and {\it vice versa}.
In a picturesque way, we can say, that only the set of the degrees
of freedom of the structure the observer belongs to is consistent
with the operation of the brain of the observer.

The arguments for  point (1)  apply to  point (2). As the only
probability density that can provide probability density for the
arbitrary subsystem of $C$ is the universal state, $\vert \Psi
\rangle$, there is not any subsystem's ($B$'s) probability
density, $\rho(X_B, X'_B)$, that could provide probability density
for {\it both} the $1$ and the $S'$ system. e.g. The definition
$X_B = f(x_1, X_{S'})$ gives rise to the probability density
$\rho(X_B, X'_B) = \rho(x_1, x'_1, X_{S'}, X'_{S'})$, which, as
emphasized above, cannot provide the probability densities
$\rho(x_1, x'_1)$ or $\rho(X_{S'}, X'_{S'})$ by integrating over
$X_{S'}$ and $x_1$, respectively. So, there is not any observable
of the  $B$ system whose measurement might approximate
simultaneous measurement of any pair of observables for the two
Brownian particles, $1$ and $S'$. Physically, this means that we
cannot imagine a third system $B$, that undergoes
Brownian-motion-like dynamics and can still approximately describe
{\it both} "microscopic" particles, $1$ and $S'$.

As we cannot recognize any other basis for emergentism, we are
forced to conclude that the above-distinguished inconsistency
between the QBM model and the modern Everett interpretation {\it
remains intact}.

\bigskip

 {\bf 4.3 Sufficiency of decoherence for branching}

\bigskip

In the absence of an emergent model-structure encompassing the
structures of Section 3, we consider and answer the following:
 Is the QBM model a proper subject of the Everett Interpretation?

To this end, we first emphasize: the standard QBM is a
(paradigmatic theoretical) {\it decoherence} model pertaining to
the {\it realistic macroscopic} situation of "Brownian motion".
Second, there are not any structural {\it phenomenological} facts
about Brownian motion that go beyond the standard QBM model
(Giulini et al 1996; Breuer and Petruccione 2002; Caldeira and
Leggett 1985; Romero and Paz 1997)--no need for any "emergent"
Brownian particle.

Bearing this in mind, the possibility that the structures
considered are not susceptible to the Everett interpretation
directly raises the following {\it foundational} question:
 Is decoherence {\it sufficient} for Everett Branching ?
If it is, then the conclusion of Section 4.2 is unavoidable. If
not, then Section 4.2 suggests an {\it additional requirement} for
branching, i.e. for completeness of the Everett interpretation is
needed. e.g. One may require some amount of "complexity" of the
composite system  to be the subject of the modern Everett
interpretation. Certainly, then the range of applicability of the
modern Everett interpretation shrinks, as distinct from the
competitive interpretations. As the "additional requirement" is
not a part of the present state of the art in the field, and is
not of interest for the  QBM model, we will not elaborate on this
any further. So we finally return to the conclusion of Section
4.2.

\bigskip

{\bf 5. Discussion}

\bigskip

Our arguments are technical. We use Entanglement Relativity and
the Parallel Occurrence of Decoherence as  corollaries of the
universally valid quantum mechanics. Thereby, every sound
interpretation must not be in conflict with them. We do not enter
any open issue of the Everett interpretation, such as the choice
of the preferred pointer basis or the origin of probability. To
this end, our considerations are non-interpretational. Essential
for our argument is the use of ER and POD together, not
separately. Nevertheless, ER+POD is not sufficient for our
argument. Actually, we apply ER+POD on a pair of very special,
mutually {\it irreducible} structures of the model universe. Only
having the irreducible structures, $1+2$ and $S'+E'$, can we argue
non-existence of an emergent Brownian particle $B$, for the pair
of Brownian particles, $1$ and $S'$, and thereby point out
inconsistency with the Everett interpretation.

In physics, "emergence" is usually linked with the "collective"
degrees of freedom of a composite system. E.g. the phonons refer
to a chain of oscillators as a whole. Knowledge about the
phonons-system state provides information (via some calculation)
about every original oscillator,  and {\it vice versa}. However,
in our considerations we ask about "emergence" for the pair $1$
and $S'$, which do not bear any structure of their own.  We show
that, as long as we require some information about $1$ and $S'$ to
be derivable from the information about $B$, then there's no room
for emergent Brownian particle $B$. If we do not require
derivability of information about $1$ and $S'$, it is not clear in
which physical sense we can claim $B$ is emergent for $1$ and
$S'$. Therefore, the two Brownian particles $1$ and $S'$ should be
considered as literally presented in Section 3--as mutually
irreducible and simultaneously evolving in time according to the
general rules of the quantum Brownian motion model (Giulini et al
1996; Breuer and Petruccione 2002; Caldeira and Leggett 1985;
Romero and Paz 1997; Dugi\' c and Jekni\' c-Dugi\' c 2012).

It appears that the only remaining way to avoid the inconsistency
with the Everett theory may be to {\it claim} physical relevance
(e.g. physical reality, in whatever sense) of only one structure,
e.g. of the $1+2$ structure, at the expense of rejecting the other
(i.e. the $S'+E'$) structure (and the related branching process)
non-physical, artificial. However, this does not seem a very
promising strategy. The only assumption of our considerations
(besides the universal validity of quantum mechanics) is that the
Universe is an {\it exactly} isolated ("closed") system, i.e. that
there is not any observer outside the Universe. Bearing this in
mind, claiming the preferred structure of the Universe does not
bear firm physical support yet. In the words of Zanardi (2001):

\smallskip

\noindent {\it "Without further physical assumption, no partition
has an ontologically superior status with respect to any other."}

\smallskip

\noindent as well as of Halliwell (2010):

\smallskip

\noindent "{\it However, for many macroscopic systems, and in
particular for the universe as a whole, there may be no natural
split into distinguished subsystems and the rest, and another way
of identifying the naturally decoherent variables is required.}"

In the absence of a clear physical rule for making a choice of the
preferred structure of the model-Universe, Section 3, we seem
obliged to return to our conclusion on insufficiency of the
Everett interpretation to properly describe the standard quantum
Brownian motion model.

\bigskip

{\bf 6. Conclusion}

\bigskip

We apply Entanglement Relativity and the parallel occurrence of
decoherence to the standard quantum Brownian motion (QBM) setup.
For the QBM model we distinguish a pair of special, mutually
irreducible structures (decompositions into subsystems) for which
neither Everett worlds branching is allowed nor can we detect any
"emergent" Brownian particle that could branch. So unless
world-branching requires additional criterion or condition, or
there is a privileged structure of the model universe, we conclude
that the universally valid and complete quantum mechanics does not
support the Everett interpretation for the QBM-model structures
considered.

\bigskip

{\bf Acknowledgements} JJD and MD acknowledge financial support
from the Ministry of Science Serbia under contract no 171028, and
partially to the COST Action MP1006.

\bigskip

{\bf REFERENCES}

\bigskip

Auyang, S. 1998 {\it Foundations of Complex-system Theories}.
Cambridge: Cambridge University Press.

Barvinsky, A. O. \& Kamenshchik, A. Yu. 1995 Preferred basis in
quantum theory and the problem of classicalization of the quantum
Universe. {\it Phys. Rev. D} {\bf 52}, 743-757.

Breuer, H.-P. \& Petruccione, F. 2002 {\it The Theory of Open
Quantum Systems}. Oxford: Clarendon Press.

Caldeira, A. O. \& Leggett, A. J. 1985 Influence of damping on
quantum interference: An exactly soluble model. {\it Phys. Rev. A}
{\bf 31}, 1059-1066.

Ciancio E., Giorda P. \& Zanardi P. 2006 Mode transformations and
entanglement relativity in bipartite Gaussian states. {\it Phys.
Lett. A} {\bf 354}, 274-280.

 De la Torre, A. C. {\it et al} 2010 Entanglement for all quantum
states. {\it Europ. J. Phys.} {\bf 31}, 325-332.

Dugi\' c  M. 1999 What is "system": the arguments from the
decoherence theory. arXiv:quant-ph/9903037v1.

Dugi\' c, M. \& Jekni\' c, J. 2006 What is "System": Some
Decoherence-Theory Arguments. {\it Int. J. Theor. Phys.} {\bf 45},
2249-2259.

 Dugi\' c, M. \& Jekni\' c-Dugi\' c, J. 2008 What Is "System":
The Information-Theoretic Arguments. {\it Int. J. Theor. Phys.}
{\bf 47}, 805-813.

Dugi\' c, M. \&  Jekni\' c-Dugi\' c, J. 2012 Parallel Occurrence
of Decoherence in Composite Quantum Systems. {\it Pramana: J.
Phys.} {\bf 79}, 199-211.

Everett, H. 1957 "Relative state" formulation of quantum
mechanics. {\it Rev. Mod. Phys.} {\bf 29}, 454-462.

Gingrich, R. M. \& Adami, C 2002 Quantum Entanglement of Moving
Bodies. {\it Phys. Rev. Lett.} {\bf 89}, 270402.

Giulini, D., Joos, E., Kiefer, C., Kupsch, J., Stamatescu, I.-O.\&
Zeh, H. D. 1996 {\it Decoherence and the Appearance of a Classical
World in Quantum Theory}. Berlin: Springer.

Halliwell, J. 2010 Emergence of Hydrodynamic Behaviour. In {\it
Many Worlds? Everett, Quantum Theory, and Reality} (ed., S.
Saunders, J. Barrett, A. Kent, \& D. Wallace), pp. 99-117. Oxford:
Oxford University Press.

Harshman, N. L. \& Wickramasekara, S. 2007 Galilean and Dynamical
Invariance of Entanglement in Particle Scattering. {\it Phys. Rev.
Lett.} {\bf 98}, 080406.

Jekni\' c-Dugi\' c, J. \& Dugi\' c, M. 2008 Multiple
System-Decomposition Method for Avoiding Quantum Decoherence. {\it
Chin. Phys. Lett.} {\bf 25}, 371-374.

Jekni\' c-Dugi\' c J. Arsenijevi\' c M, Dugi\' c M., 2013 {\it
Quantum Structures: A View of a Quantum World}. Saarbrucken:
Lambert Academic Publishing; also available as arXiv:1306.5471
[quant-ph].

Kent, A. 2012  Real World Interpretations of Quantum Theory. {\it
Found. Phys.} {\bf 42}, 421-435.

Mc Weeney R. 1978 {\it Methods of Molecular Quantum Mechanics}.
New York: Academic Press

Romero, L. D. \& Paz, J. P. 1997 Decoherence and initial
correlations in quantum Brownian motion. {\it Phys. Rev. A} {\bf
55}, 4070-4083.

Saunders, S. 2010 Many Worlds? An Introduction. In {\it Many
Worlds? Everett, Quantum Theory, and Reality} (ed., S. Saunders,
J. Barrett, A. Kent, \& D. Wallace), pp. 1-52. Oxford: Oxford
University Press.

Terra Cunha, M. O., Dunningham, J. A. \&   Vedral, V. 2007
Entanglement in single-particle systems. {\it Proc. R. Soc. A}
{\bf 463}, 2277-2286.

Tommasini, P., Timmermans, E. \& Piza, A.F.R.D. 1998 The hydrogen
atom as an entangled electron-proton system. {\it Am. J. Phys.}
{\bf 66}, 881-885.

Valentini, A. 2010 De Broglie-Bohm Pilot-Wave Theory: Many Worlds
in Denial?. In {\it Many Worlds? Everett, Quantum Theory, and
Reality} (ed., S. Saunders, J. Barrett, A. Kent, \& D. Wallace),
pp. 476-509. Oxford: Oxford University Press.

von Neumann, J. 1955 {\it Mathematical foundations of quantum
mechanics}. Princeton: Princeton University Press.

Wallace, D. 2010 Decoherence and Ontology. In {\it Many Worlds?
Everett, Quantum Theory, and Reality} (ed., S. Saunders, J.
Barrett, A. Kent \& D. Wallace), pp. 53-72. Oxford: Oxford
University Press.

Wallace, D. 2012 {\it The Emergent Multiverse: Quantum Theory
according to the Everett Interpretation}. Oxford: Oxford
University Press.

Zanardi, P. 2001 Virtual Quantum Subsystems. {\it Phys. Rev.
Lett.} {\bf 87}, 077901.

\end{document}